\documentclass{ws-ijmpa}
\usepackage[super,compress]{cite}
\usepackage{graphicx}
\usepackage{xcolor}
\begin{document}
\markboth{Sebasti\'an Franco}{2d (0,2) Gauge Theories from Branes: Recent Progress in Brane Brick Models}

\catchline{}{}{}{}{}
%
%%%%%%%%%%%%%%%%%%%%%%%%%%%%%%%%%%%%%%%%%%%%%%%%%%%%%%%%%%%%%%%%%%%%

\title{2d (0,2) Gauge Theories from Branes: \\
Recent Progress in Brane Brick Models}

\author{Sebasti\'an Franco}

\address{Physics Department, The City College of the CUNY\\
	160 Convent Avenue, New York, NY 10031, USA \\
Physics Program and Initiative for the Theoretical Sciences\\
	The Graduate School and University Center, The City University of New York\\
	365 Fifth Avenue, New York NY 10016, USA \\
	sfranco@ccny.cuny.edu}

\maketitle

%\begin{history}
%\received{Day Month Year}
%\revised{Day Month Year}
%\end{history}

\begin{abstract}
We discuss the realization of $2d$ $(0,2)$ gauge theories in terms of branes focusing on Brane Brick Models, which are T-dual to D1-branes probing toric Calabi-Yau 4-folds. These brane setups fully encode the infinite class of $2d$ $(0,2)$ quiver gauge theories on the worldvolume of the D1-branes and substantially streamline their connection to the probed geometries. We review various methods for efficiently generating Brane Brick Models. These algorithms are then used to construct $2d$ $(0,2)$ gauge theories for the cones over all the smooth Fano 3-folds and two infinite families of Sasaki-Einstein 7-manifolds with known metrics. This note is based on the author’s talk at the Gauged Linear Sigma Models @ 30 conference at the Simons Center for Geometry and Physics.

\end{abstract}

\ccode{PACS numbers:}

%\tableofcontents

%===================================================
\section{Introduction}	
%===================================================

Engineering quantum field theories in terms of String or M-Theory branes is a powerful approach for studying their dynamics, often providing alternative perspectives. D-branes probing singularities provide a platform for constructing interesting field theories in different dimensions. $4d$ $\mathcal{N}=1$ gauge theories on D3-branes probing singular toric Calabi-Yau (CY) 3-folds are the most thoroughly studied setups within this class. In this case, {\it brane tilings} significantly streamline the connection between the $4d$ gauge theories on the D3-branes and the probed geometry \cite{Franco:2005rj,Franco:2005sm}. 

In recent years, the study of $2d$ $(0,2)$ gauge theories on D1-branes probing toric CY 4-folds culminated with the introduction of {\it brane brick models} \cite{Franco:2015tya}, which result in powerful simplifications in the map between geometry and $2d$ theories analogous to the ones previously brought by brane tilings. 

In this note, we present a brief review of recent works in which brane brick models have been exploited for determining and studying the gauge theories for large classes of interesting CY 4-folds. These geometries include the complex cones over all smooth Fano 3-folds \cite{Franco:2022gvl}, and infinite families of cones over the $Y^{p,k}(\mathbb{CP}^{1}\times \mathbb{CP}^{1})$ and $Y^{p,k}(\mathbb{CP}^{2})$ Sasaki-Einstein 7-manifolds \cite{Franco:2022isw}.

%===================================================
\section{Brane Brick Models}	
%===================================================

\label{section_BBMs}

Brane brick models are obtained from D1-branes at $\text{CY}_4$ singularities by T-duality. We refer the reader to \cite{Franco:2015tna,Franco:2015tya,Franco:2016nwv,Franco:2016qxh} for detailed discussions. A brane brick model is a Type IIA brane configuration consisting of D4-branes wrapping a 3-torus $\mathbb{T}^3$ and suspended from an NS5-brane that wraps a holomorphic surface $\Sigma$ intersecting with $\mathbb{T}^3$. The holomorphic surface $\Sigma$ is the zero locus of the Newton polynomial defined by the toric diagram of the $\text{CY}_4$. The basic ingredients of the brane setup are summarized in Table \ref{Brane brick-config}. The $(246)$ directions are compactified on a $\mathbb{T}^3$. The $2d$ gauge theory lives on the two directions $(01)$ common to all the branes.

%======================================================================
\begin{table}[ph]
\tbl{Brane brick model configuration.}
{\begin{tabular}{l|cccccccccc}
\; & 0 & 1 & 2 & 3 & 4 & 5 & 6 & 7 & 8 & 9 \\
\hline
$\text{D4}$ & $\times$ & $\times$ & $\times$ & $\cdot$ & $\times$ & $\cdot$ & $\times$ & $\cdot$ & $\cdot$ & $\cdot$  \\
$\text{NS5}$ & $\times$ & $\times$ & \multicolumn{6}{c}{----------- \ $\Sigma$ \ ------------} & $\cdot$ & $\cdot$ \\
\end{tabular} 
\label{Brane brick-config}}
\end{table}
%======================================================================

Brane brick models, or equivalently their dual periodic quivers, fully encode the $2d$ $(0,2)$ quiver gauge theories on the worldvolume of D1-branes probing toric CY 4-folds. Namely, they summarize not only the quivers but also the $J$- and $E$-terms. The dictionary between brane brick models and gauge theories is summarized in Table \ref{tbrick}.

%======================================================================
\begin{table}[ph]
\tbl{Dictionary between brane brick models and $2d$ $(0,2)$ gauge theories.}
{\begin{tabular}{|c|c|c|}
\hline
{\bf Brane Brick Model} \ \ &  {\bf Gauge Theory} \ \ \ \ \ \ \  & {\bf Periodic Quiver} \ \ \ 
\\
\hline\hline
Brick  & Gauge group & Node \\
\hline
Oriented face  & Bifundamental chiral field & Oriented (black) arrow 
\\
between bricks $i$ and $j$ & from node $i$ to node $j$  & from node $i$ to node $j$ \\
\hline
Unoriented square face  & Bifundamental Fermi field & Unoriented (red) line \\
between bricks $i$ and $j$ & between nodes $i$ and $j$ & between nodes $i$ and $j$  \\
\hline
Edge  & $J$- or $E$-term & Plaquette encoding \\ 
& & a $J$- or an $E$-term \\
\hline
\end{tabular}
\label{tbrick}}
\end{table}
%======================================================================

For additional results regarding brane brick models, we refer the interested reader to \cite{Franco:2017cjj,Franco:2021elb,Franco:2023tyf}.

%===================================================
\section{From CY$_4$'s to Brane Brick Models}	
%===================================================

\label{section_methods}

\begin{figure}[h]
	\centering
	\includegraphics[height=8cm]{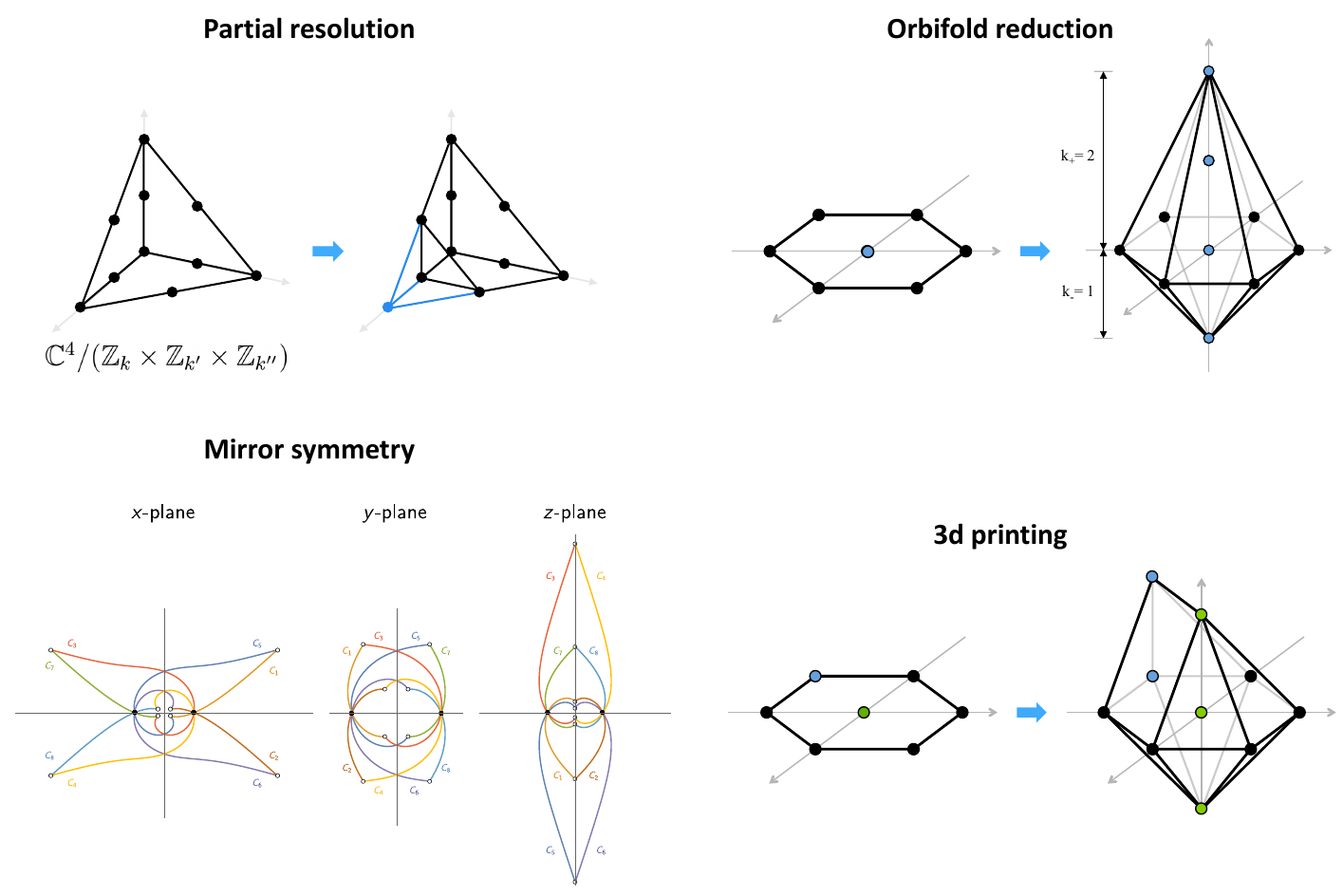}
\caption{Some methods for generating brane brick models for toric CY 4-folds.}
	\label{methods_construction_BBMs}
\end{figure}

Several methods for constructing brane brick models associated to a given toric CY$_4$ have been developed. Some of them considerably simplify this task. Figure \ref{methods_construction_BBMs} summarizes a few of these procedures. 

{\it Partial resolution} consists of embedding the toric diagram of interest within a larger toric diagram, for which the $2d$ gauge theory is known. A standard choice for such initial geometry is an abelian orbifold of $\mathbb{C}^4$. The deletion of points that connects the two toric diagrams translates into higgsing in the field theory \cite{Franco:2015tna}. The determination of the chiral fields that acquire a non-zero VEV to achieve a desired toric diagram is simplified by considering the map between fields in the quiver and {\it brick matchings}, certain combinatorial objects in the associated brane brick model that are analogous to perfect matchings of brane tilings \cite{Franco:2015tya}. 

{\it Orbifold reduction} generates the $2d$ $(0, 2)$ gauge theories associated to D1-branes probing a toric CY$_4$ starting from $4d$ $\mathcal{N} = 1$ gauge theories on D3-branes probing toric CY$_3$’s or, equivalently, the corresponding brane tilings. Given two integers $k_+,k_- \geq 0$ and a perfect matching $p_0$ of a brane tilings for a CY$_3$, orbifold reduction generates a gauge theory that corresponds to a CY$_4$ whose toric diagram is obtained by expanding the point associated to $p_0$ into a line of length $k_+ + k_-$, with $k_+$ points above the original $2d$ toric diagram and $k_-$points below it. The algorithm for producing the $2d$ gauge theory has an elegant implementation in terms of the periodic quiver \cite{Franco:2016fxm}. This procedure generalizes dimensional reduction and orbifolding. With orbifold reduction, the gauge theories for rather complicated CY$_4$’s can be found with little effort.

{\it $3d$ printing} is similar to orbifold reduction in that it provides a combinatorial prescription for constructing the $2d$ (0,2) gauge theory for a CY$_4$ starting from the brane tiling for a CY$_3$ \cite{Franco:2018qsc}. In $3d$ printing, multiple points in the toric diagram of the CY$_3$ can be lifted to produce the toric diagram of the CY$_4$, as illustrated in Figure \ref{methods_construction_BBMs}. 

{\it Mirror symmetry} provides an alternative way for deriving the $2d$ $(0,2)$ gauge theory. The mirror configuration consists of D5-branes wrapping 4-spheres and the gauge theory is determined by how they intersect \cite{Franco:2016qxh,Futaki:2014mpa}. Figure \ref{methods_construction_BBMs}, shows various projections that contribute to the visualization of the configuration of branes in the mirror. Interestingly, changing the complex structure and passing through vanishing cycles results in inequivalent geometries. However, the mirror geometry unifies the inequivalent geometries of the CY into a single CY manifold. In this way we can understand field theory dualities from the uniqueness of the CY mirror.  

%===================================================================
%===================================================================

%===================================================
\section{Fano 3-Folds}	
%===================================================

Brane brick models are particularly useful for finding the $2d$ (0,2) gauge theories for large families of toric CY 4-folds. An interesting family of geometries is given by the complex cones over Gorenstein Fano varieties that are constructed from a special set of lattice polytopes known as reflexive polytopes. Table \ref{t_counting} summarizes the numbers of inequivalent reflexive polytopes up to dimension 4, following the seminal classification of Kreuzer and Skarke \cite{Kreuzer:1995cd,Kreuzer:1998vb,Kreuzer:2000xy}. 

%======================================================================
\begin{table}[h]
\tbl{The number of inequivalent reflexive polytopes and regular reflexive polytopes in dimension $d\leq 4$ \cite{Kreuzer:1995cd,Kreuzer:1998vb,Kreuzer:2000xy}.}
{\begin{tabular}{|c|c|c|}
\hline
$d$ & Number of Polytopes & Number of Regular Polytopes \\
\hline \hline
1 & 1  & 1\\
\hline
2 & 16 & 5\\
\hline
3 & 4,319 & 18 \\
\hline 
4 & 473,800,776 & 124 \\
\hline
\end{tabular}
\label{t_counting}}
\end{table}
%======================================================================

\begin{figure}[h]
	\centering
	\includegraphics[height=10cm]{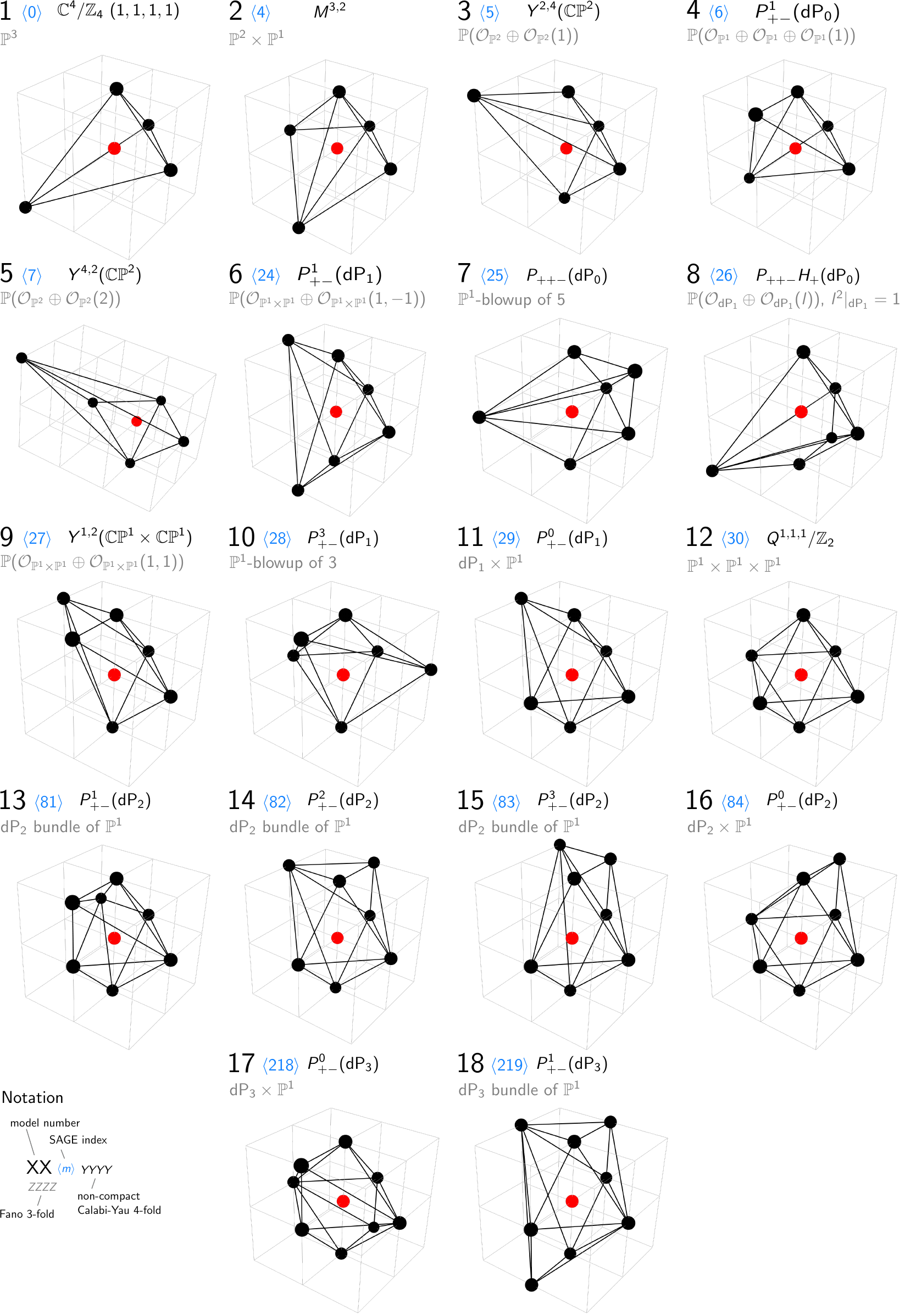}
\caption{The 18 regular reflexive polytopes in dimension 3 corresponding to toric non-compact CY 4-folds and corresponding smooth Fano 3-folds.}
	\label{f_alltoricdia}
\end{figure}

These polytopes should be regarded as the toric diagrams of toric CY's constructed as the complex cones over the corresponding Fanos. We are interested in CY 4-folds, i.e. dimension 3, where there are 4,319 polytopes. Moreover, let us focus on those reflexive polytopes that are also regular, which implies that their associated Gorenstein Fano varieties are smooth. According to Table \ref{f_alltoricdia}, this leaves us with a manageable subset of 18 polytopes, shown in Figure \ref{f_alltoricdia}. 

The CY$_3$ analogues of these geometries are the complex cones over $F_0$ and del Pezzo surfaces, which have played a prominent role in elucidating the correspondence between CY 3-folds and the $4d$ $\mathcal{N}=1$ gauge theories on D3-branes probing them (see e.g. \cite{Feng:2000mi,Feng:2001xr,Beasley:2001zp,Feng:2002zw}). 

%===================================================================
%===================================================================

In \cite{Franco:2022gvl}, a brane brick model, i.e. a $2d$ $(0,2)$ gauge theory, was constructed for each of the 18 regular reflexive polytopes in 3 dimensions. For every one of these models, the moduli space was thoroughly studied, calculating the generating function of mesonic gauge invariant operators, the Hilbert series, using the Molien integral formula. For each of these models, the generators of the mesonic moduli space were expressed both in terms of chiral fields of the $2d$ gauge theory as well as brick matchings. Finally, for all these models, it was verified that the generator lattice of the corresponding mesonic moduli space is the polar reflexive dual of the toric diagram.

%===================================================
\subsection{An Example: Model 11}	
%===================================================

\begin{figure}[h]
	\centering
	\includegraphics[height=3cm]{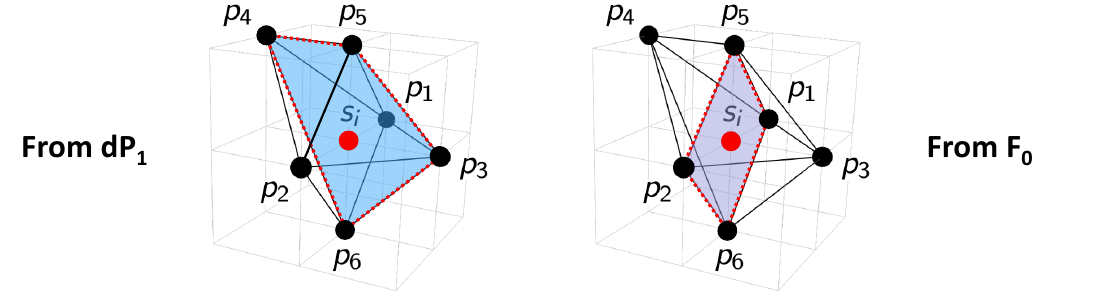}
\caption{Two alternative ways of obtaining the toric diagram for Model 11 from CY$_3$ toric diagrams.}
	\label{toric_diagrams_Model_11}
\end{figure}

To illustrate the results in \cite{Franco:2022gvl}, let us focus on Model 11. Figure \ref{toric_diagrams_Model_11} shows how the toric diagram for this model can be connected to two different CY$_3$ toric diagrams. Therefore, it is possible to derive the corresponding $2d$ $(0,2)$ gauge theory using some of the approaches reviewed in Section \ref{section_methods}. Starting from $dP_1$, the toric diagram for Model 11 is obtained by lifting the central point in two opposite directions. The associated gauge theory can be therefore constructed from the one for $dP_1$ using orbifold reduction. Alternatively, the toric diagram for Model 11 follows from lifting two points in the toric diagram of $F_0$. Consequently, the gauge theory for Model 11 can also be constructed by $3d$ printing starting from the gauge theory for $F_0$. It is interesting to reflect on how the $SU(2)\times SU(2)$ global symmetry of the final gauge theory arises from these two alternative constructions. In the first one, only one of the $SU(2)$ factors is present in $dP_1$, while the second one emerges from orbifold reduction. In contrast, the full $SU(2)\times SU(2)$ symmetry is already present in $F_0$.

%===================================================================
%===================================================================

%===================================================================
\begin{figure}[h]
	\centering
	\includegraphics[height=4.5cm]{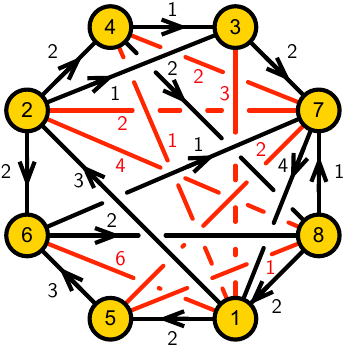}
\caption{Quiver for Model 11.}
	\label{quiver_11}
\end{figure}
%===================================================================

The corresponding brane brick model has the quiver in Figure \ref{quiver_11} and the $J$- and $E$-terms are 
\begin{equation}
{\footnotesize
\begin{array}{rrclrcl}
 &  & J &  & & E &  \\
\Lambda_{16}^{1}  : & Y_{68} X_{81}  &-& X_{67} X_{71}  &  P_{15} X_{56}  &-& X_{12} P_{26} \\ 
\Lambda_{16}^{2}  : & X_{67} Y_{71}  &-& X_{68} X_{81} &  P_{15} Y_{56}  &-& Y_{12} P_{26} \\ 
\Lambda_{16}^{3}  : & X_{68} X_{87} X_{71}  &-& Y_{68} X_{87} Y_{71} &  P_{15} Z_{56}  &-& Z_{12} P_{26} \\ 
\Lambda_{16}^{4}  : & Y_{68} S_{81}  &-& X_{67} S_{71} &  X_{12} Q_{26}  &-& Q_{15} X_{56} \\ 
\Lambda_{16}^{5}  : & X_{67} T_{71}  &-& X_{68} S_{81} &  Y_{12} Q_{26}  &-& Q_{15} Y_{56} \\ 
\Lambda_{16}^{6}  : & X_{68} X_{87} S_{71}  &-& Y_{68} X_{87} T_{71} &  Z_{12} Q_{26}  &-& Q_{15} Z_{56} \\ 
\Lambda_{27}^{1}  : & Y_{71} Y_{12}  &-& X_{71} X_{12}  &  P_{26} X_{67}  &-& X_{23} P_{37} \\ 
\Lambda_{27}^{2}  : & T_{71} Y_{12}  &-& S_{71} X_{12}  &  X_{23} Q_{37}  &-& Q_{26} X_{67} \\ 
\Lambda_{28}^{1}  : & X_{87} X_{71} Z_{12}  &-& X_{81} Y_{12} &  P_{26} X_{68}  &-& X_{24} P_{48} \\ 
\Lambda_{28}^{2}  : & X_{81} X_{12}  &-& X_{87} Y_{71} Z_{12} &  P_{26} Y_{68}  &-& Y_{24} P_{48} \\ 
\Lambda_{28}^{3}  : & X_{87} S_{71} Z_{12}  &-& S_{81} Y_{12}  &  X_{24} Q_{48}  &-& Q_{26} X_{68} \\ 
\Lambda_{28}^{4}  : & S_{81} X_{12}  &-& X_{87} T_{71} Z_{12} &  Y_{24} Q_{48}  &-& Q_{26} Y_{68}  \\ 
\Lambda_{31}^{1}  : & Z_{12} X_{24} X_{43}  &-& X_{12} X_{23}  &  P_{37} X_{71}  &-& Q_{37} S_{71} \\ 
\Lambda_{31}^{2}  : & Y_{12} X_{23}  &-& Z_{12} Y_{24} X_{43} &  P_{37} Y_{71}  &-& Q_{37} T_{71} \\ 
\Lambda_{41}  : & X_{12} Y_{24}  &-& Y_{12} X_{24} &  P_{48} X_{81}  &-& Q_{48} S_{81} \\ 
\Lambda_{47}^{1}  : & X_{71} Z_{12} X_{24}  &-& Y_{71} Z_{12} Y_{24} &  P_{48} X_{87}  &-& X_{43} P_{37} \\ 
\Lambda_{47}^{2}  : & S_{71} Z_{12} X_{24}  &-& T_{71} Z_{12} Y_{24} &  X_{43} Q_{37}  &-& Q_{48} X_{87} \\ 
\Lambda_{75}^{1}  : & Z_{56} X_{68} X_{87}  &-& X_{56} X_{67}  &  S_{71} Q_{15}  &-& X_{71} P_{15} \\ 
\Lambda_{75}^{2}  : & Y_{56} X_{67}  &-& Z_{56} Y_{68} X_{87} &  T_{71} Q_{15}  &-& Y_{71} P_{15} \\ 
\Lambda_{85}  : & X_{56} Y_{68}  &-& Y_{56} X_{68} &  S_{81} Q_{15}  &-& X_{81} P_{15}
 \end{array} 
 }
\label{E_J_C_+-}
\end{equation}

\begin{figure}[h]
	\centering
	\includegraphics[height=4.5cm]{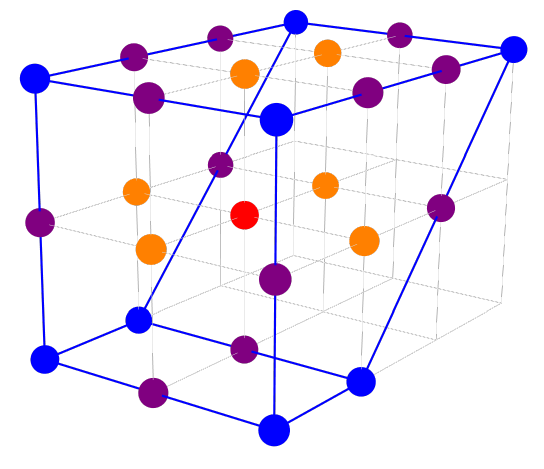}
\caption{Generator lattice for Model 11.}
	\label{genlattice_11}
\end{figure}

Table \ref{table_Model_11} presents the generators of the mesonic moduli space of Model 11 in terms of brick matchings with the corresponding flavor charges. Figure \ref{genlattice_11} shows the corresponding generator lattice, which is a reflexive polytope that is the dual of the toric diagram of Model 11 shown in Figure \ref{f_alltoricdia}, as expected. 

%===================================================
\begin{table}[h!]
\centering
\tbl{The generators of the mesonic moduli space of Model 11 in terms of brick matchings with the corresponding flavor charges.}
{\begin{tabular}{|c|c|c|c|}
\hline
Generator & $SU(2)_{\tilde{x}}$ & $SU(2)_{\tilde{y}}$ & $U(1)_{\tilde{b}}$ \\
\hline
$p_1^2 p_3 p_6^2 ~s o  $ &1& 0& -1\\
$p_1 p_2 p_3 p_6^2 ~s o  $ &0& 0& -1\\
$p_2^2 p_3 p_6^2 ~s o  $ &-1& 0& -1\\
$p_1^2 p_4 p_6^2 ~s o  $ &1& -1& -1\\
$p_1 p_2 p_4 p_6^2 ~s o  $ &0& -1& -1\\
$p_2^2 p_4 p_6^2 ~s o  $ &-1& -1& -1\\
$p_1^2 p_3^2 p_5 p_6 ~s o ^2 $ &1& 1& 0\\
$p_1 p_2 p_3^2 p_5 p_6 ~s o ^2 $ &0& 1& 0\\
$p_2^2 p_3^2 p_5 p_6 ~s o ^2 $ &-1& 1& 0\\
$p_1^2 p_3 p_4 p_5 p_6 ~s o ^2 $ &1& 0& 0\\
$p_1 p_2 p_3 p_4 p_5 p_6 ~s o ^2 $ &0& 0& 0\\
$p_2^2 p_3 p_4 p_5 p_6 ~s o ^2 $ &-1& 0& 0\\
$p_1^2 p_4^2 p_5 p_6 ~s o ^2 $ &1& -1& 0\\
$p_1 p_2 p_4^2 p_5 p_6 ~s o ^2 $ &0& -1& 0\\
$p_2^2 p_4^2 p_5 p_6 ~s o ^2 $ &-1& -1& 0\\
$p_1^2 p_3^3 p_5^2 ~s o ^3 $ &1& 2& 1\\
$p_1 p_2 p_3^3 p_5^2 ~s o ^3 $ &0& 2& 1\\
$p_2^2 p_3^3 p_5^2 ~s o ^3 $ &-1& 2& 1\\
$p_1^2 p_3^2 p_4 p_5^2 ~s o ^3 $ &1& 1& 1\\
$p_1 p_2 p_3^2 p_4 p_5^2 ~s o ^3 $ &0& 1& 1\\
$p_2^2 p_3^2 p_4 p_5^2 ~s o ^3 $ &-1& 1& 1\\
$p_1^2 p_3 p_4^2 p_5^2 ~s o ^3 $ &1& 0& 1\\
$p_1 p_2 p_3 p_4^2 p_5^2 ~s o ^3 $ &0& 0& 1\\
$p_2^2 p_3 p_4^2 p_5^2 ~s o ^3 $ &-1& 0& 1\\
$p_1^2 p_4^3 p_5^2 ~s o ^3 $ &1& -1& 1\\
$p_1 p_2 p_4^3 p_5^2 ~s o ^3 $ &0& -1& 1\\
$p_2^2 p_4^3 p_5^2 ~s o ^3 $ &-1& -1& 1\\
\hline
\end{tabular}}
\label{table_Model_11}
\end{table}
%===================================================

%===================================================================
%===================================================================

In \cite{Franco:2022gvl}, all generators were also expressed in terms of chiral fields in the quiver. As an example, Table \ref{f_genfields_11a} provides these expressions for the first two generators in Table \ref{table_Model_11}. Every generator can be represented in multiple ways in terms of the fields in the gauge theory.

%===================================================
\begin{table}[h]
\centering
\tbl{First two generators in Table \ref{table_Model_11} expressed in terms of chiral fields in the quiver.}
{\begin{tabular}{|c|c|c|c|}
\hline
Generator & $SU(2)_{\tilde{x}}$ & $SU(2)_{\tilde{y}}$ & $U(1)_{\tilde{b}}$ \\
\hline
$P_{15} Z_{56} X_{67} S_{71}=  P_{15} Z_{56} Y_{68} S_{81}=  Z_{12} P_{26} X_{67} S_{71}=$ &1& 0& -1\\ 
$=Z_{12} P_{26} Y_{68} S_{81}=  Z_{12} X_{23} P_{37} S_{71}=  Z_{12} Y_{24} P_{48} S_{81}$ & & & \\ \hline
$P_{15} Z_{56} X_{67} X_{71}=  P_{15} Z_{56} Y_{68} X_{81}=  Z_{12} P_{26} X_{67} X_{71}= $  &0& 0& -1\\
$=Z_{12} P_{26} Y_{68} X_{81}=  Z_{12} X_{23} P_{37} X_{71}=  Z_{12} Y_{24} P_{48} X_{81}$ & & & \\ \hline
\end{tabular}}
\label{f_genfields_11a}
\end{table}
%===================================================

%===================================================
\section{Sasaki-Einstein 7-Manifolds}	
%===================================================

Every $2n$-dimensional K\"ahler-Einstein manifold $B_{2n}$ there is an infinite family of compact Sasaki-Einstein (SE) manifolds $Y_{2n+3}$ of dimension $2n+3$ \cite{Gauntlett:2004hh,Martelli:2008rt}. 
For $n=2$, the 4-dimensional K\"ahler-Einstein bases $B_4$ are either $\mathbb{CP}^{1}\times \mathbb{CP}^{1}$ or $\mathbb{CP}^{2}$, giving rise to two infinite families of SE 7-manifolds denoted $Y^{p,k}(\mathbb{CP}^{1}\times\mathbb{CP}^{1})$ and $Y^{p,k}(\mathbb{CP}^{2})$, respectively. The two families stand out because their SE metrics are known explicitly \cite{Gauntlett:2004hh,Martelli:2008rt}. 

\begin{figure}[h]
	\centering
	\includegraphics[height=5.5cm]{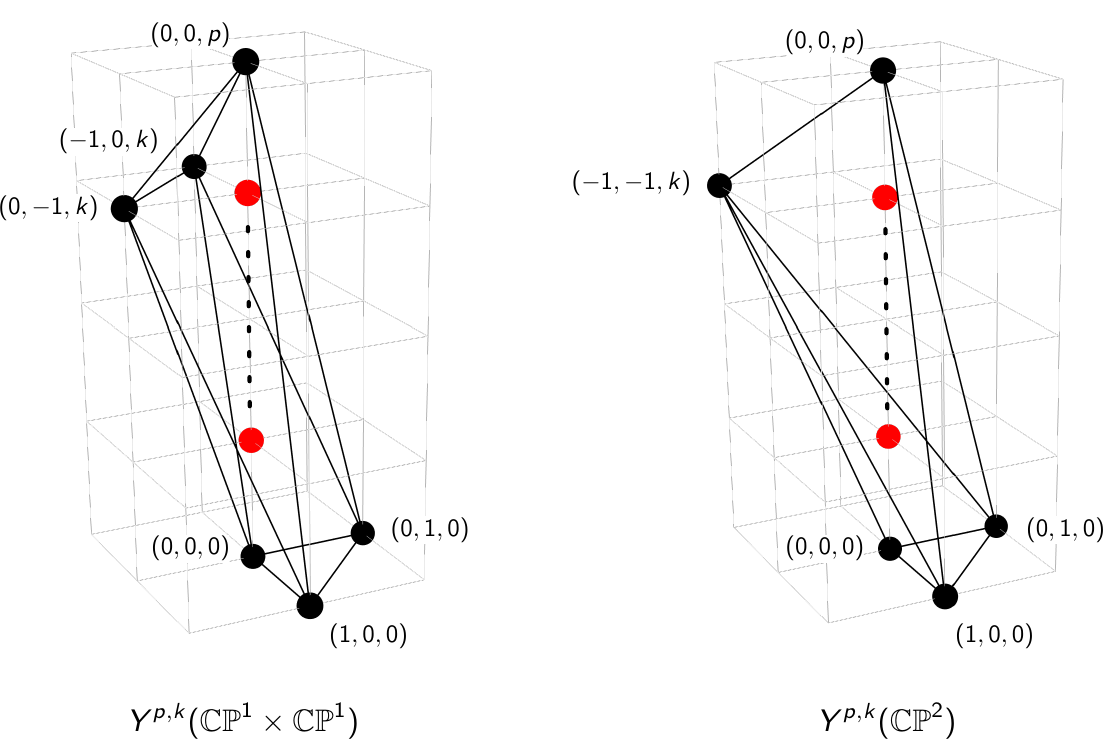}
\caption{General toric diagrams for the $Y^{p,k}(\mathbb{CP}^2)$ and $Y^{p,k}(\mathbb{CP}^1\times \mathbb{CP}^1)$ families of toric CY 4-folds.}
	\label{toric_ypk}
\end{figure}

The general toric diagrams for the corresponding CY 4-folds are show in Figure \ref{toric_ypk}\footnote{For brevity, we use the name of the SE base to also identify the corresponding CY$_4$.}, where the ranges for the parameters $p$ and $k$ are 
\begin{eqnarray}
Y^{p,k}(\mathbb{CP}^{1}\times\mathbb{CP}^{1}) &~:~ 
0 \leq k \leq p 
~,~
\nonumber \\
Y^{p,k}(\mathbb{CP}^{2}) &~:~ 
0 \leq k \leq \frac{3}{2} p 
~.~
\end{eqnarray}

%===================================================================
%===================================================================

The isometry of the SE 7-manifolds takes the general form $H \times U(1)^2$, with $H$ the isometry of the base $B_4$. 
For $Y^{p,k}(\mathbb{CP}^{1}\times\mathbb{CP}^{1})$ and $Y^{p,k}(\mathbb{CP}^{2})$, the isometries are $SU(2)\times SU(2) \times U(1)^2$ and $SU(3)\times U(1)^2$, respectively.  These isometries translate into the global symmetries of the corresponding $2d$ $(0,2)$ gauge theories. 

\begin{figure}[h]
	\centering
	\includegraphics[height=6cm]{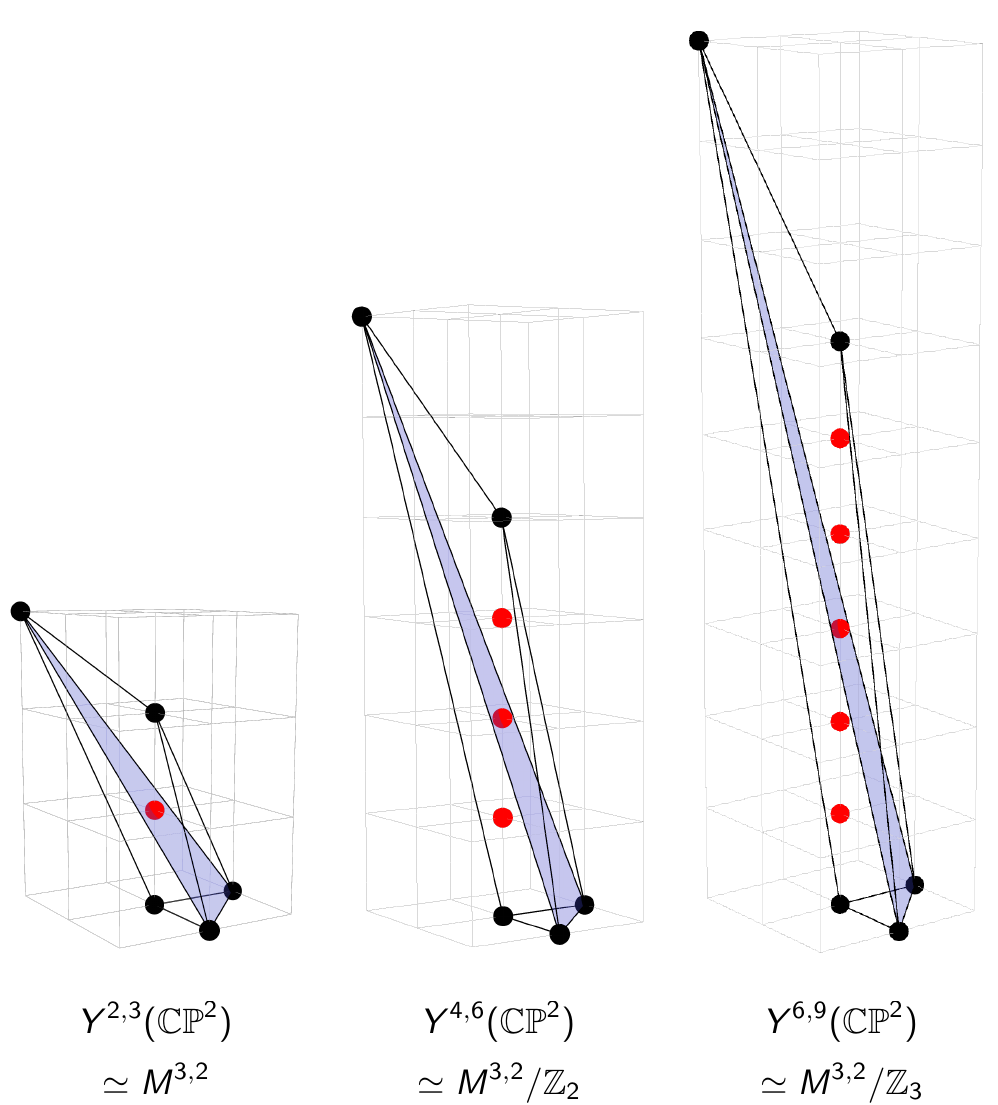}
\caption{Various $Y^{p,k}(\mathbb{CP}^{2})$ geometries that can be obtained via orbifold reduction from $dP_0$.}
	\label{yppcp2_orbifold_reduction}
\end{figure}

The $2d$ $(0,2)$ gauge theories for both infinite families of CY 4-folds were constructed in \cite{Franco:2022isw}, guided by the techniques presented in Section \ref{section_methods}. For example, the cones over $Y^{p,k}(\mathbb{CP}^{2})$ with $p=\frac{2}{3} k = 2m$ and $m\in\mathbb{Z}^{+}$ are equivalent to abelian orbifolds of the form $M^{3,2}/\mathbb{Z}_m$. The corresponding $2d$ $(0,2)$ theories can be obtained via orbifold reduction of the $4d$ $\mathcal{N}=1$ theory corresponding to $dP_0$. This is made clear by Figure \ref{yppcp2_orbifold_reduction}, where the highlighted plane indicates the toric diagram for $dP_0$. While not all the $Y^{p,k}(\mathbb{CP}^{2})$ theories can be constructed in this way, the models we obtain contain sufficient information to propose a closed form for the entire family, a proposal that can then be checked to be correct. The gauge theories for the full $Y^{p,k}(\mathbb{CP}^{1}\times\mathbb{CP}^{1})$ family can be determined using the same approach.

%===================================================================
%===================================================================

\begin{figure}[h]
	\centering
	\includegraphics[height=6cm]{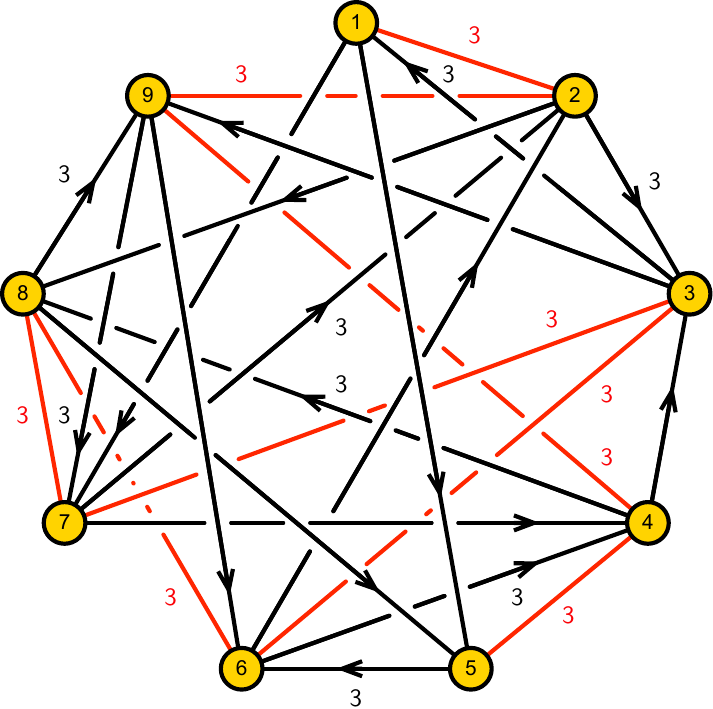}
\caption{Quiver for $Y^{3,1}(\mathbb{CP}^2)$.}
	\label{f_quiver_cp2_31}
\end{figure}

As an example, the quiver for the $2d$ $(0,2)$ theory corresponding to $Y^{3,1}(\mathbb{CP}^2)$ is shown in Figure \ref{f_quiver_cp2_31}. The corresponding $J$- and $E$-terms take the following form
\begin{equation}
\begin{array}{rccrcl}
 &   J &   & & E &  \\
 \Lambda_{12}^{i}  : & \epsilon_{ijk} X^{j}_{23} X^{k}_{31 } &~&  P_{15} X^{i}_{56} Q_{62 } &-&  Q_{17} X^{i}_{72 } \\
 \Lambda_{29}^{2i}  : & \epsilon_{ijk} X^{j}_{97} X^{k}_{72 } &~&  X^{i}_{23} Q_{39 } &-&  Q_{28} X^{i}_{89 } \\
 \Lambda_{37}^{2i}  : & \epsilon_{ijk}  X^{j}_{72} X^{k}_{23 } &~&  X^{i}_{31} Q_{17 } &-&  Q_{39} X^{i}_{97} \\
 \Lambda_{45}^{i}  : & \epsilon_{ijk} X^{j}_{56} X^{k}_{64 } &~&  X^{i}_{48} Q_{85 } &-&   Q_{43} X^{i}_{31} P_{15} \\
 \Lambda_{63}^{2i}  : & \epsilon_{ijk} X^{j}_{31} P_{15} X^{k}_{56 } &~&  X^{i}_{64} Q_{43 } &-&  Q_{62} X^{i}_{23} \\
 \Lambda_{78}^{i}  : & \epsilon_{ijk} X^{j}_{89} X^{k}_{97 } &~&  X^{i}_{72} Q_{28 } &-&  Q_{74} X^{i}_{48 } \\
 \Lambda_{86}^{2i}  : & \epsilon_{ijk} X^{j}_{64} X^{k}_{48 } &~&  X^{i}_{89} Q_{96 } &-&  Q_{85} X^{i}_{56} \\
 \Lambda_{94}^{2i}  : & \epsilon_{ijk} X^{j}_{48} X^{k}_{89 } &~&  X^{i}_{97} Q_{74 } &-&  Q_{96} X^{i}_{64} \\
 \end{array} 
~,~
\label{es0500}
\end{equation}
where $i,j,k=1,2,3$ are $SU(3)$ global symmetry indices.

%===================================================================
%===================================================================

\begin{figure}[h]
	\centering
	\includegraphics[height=4cm]{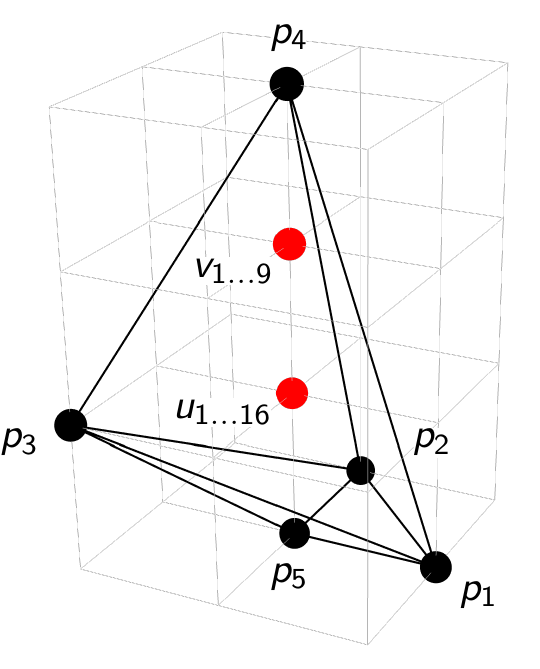}
\caption{Toric diagram of $Y^{3,1}(\mathbb{CP}^2)$.}
	\label{f_toric_cp2_31}
\end{figure}

Computing the moduli space of this theory, one obtains the toric diagram in Figure \ref{f_toric_cp2_31}, which is indeed the one for $Y^{3,1}(\mathbb{CP}^2)$.

%===================================================================
%===================================================================

%===================================================
\section*{Acknowledgments}
%===================================================

It is a pleasure to thank the organizers of Gauged Linear Sigma Models @ 30 for putting together such an exciting meeting and for the opportunity to present my work. I would also like to acknowledge the staff at the Simons Center for Geometry and Physics for their wonderful hospitality. I am grateful to Dongwook Ghim and Rak-Kyeong Seong for enjoyable collaborations on the main works discussed in this presentation. This work is supported by the U.S. National Science Foundation grants PHY-2112729 and DMS-1854179.

%%%%%%%%%%%%%%%%%%%%%%%%%%%%%%%%%%%%%%%%%%%%%%%%%%%%%%%%%%%%%%%%%%%%%%%%%

%\begin{thebibliography}{000} %for 3 digits
%\begin{thebibliography}{00}  %for 2 digits

\end{document}